# Optical Mode Control by Geometric Phase in Quasicrystal Metasurface

Igor Yulevich, Elhanan Maguid, Nir Shitrit, Dekel Veksler, Vladimir Kleiner, and Erez Hasman[*]

*Micro and Nanooptics Laboratory, Faculty of Mechanical Engineering, and Russell Berrie Nanotechnology Institute, Technion – Israel Institute of Technology, Haifa 32000, Israel*

[*]e-mail: mehasman@technion.ac.il

**Abstract**

We report on the observation of optical spin-controlled modes from a quasicrystalline metasurface as a result of an aperiodic geometric phase induced by anisotropic subwavelength structure. When geometric phase defects are introduced in the aperiodic structured surface, the modes exhibit polarization helicity dependence resulting in the optical spin-Hall effect. The radiative thermal dispersion bands from a quasicrystal structure were studied where the observed bands arise from the optical spin-orbit interaction induced by the aperiodic space-variant orientations of anisotropic antennas. The optical spin-flip behavior of the revealed modes that arise from the geometric phase pickup was experimentally observed within the visible spectrum by measuring the spin-projected diffraction patterns. The introduced ability to manipulate the light-matter interaction of quasicrystals in a spin-dependent manner provides the route for molding light via spin-optical aperiodic artificial planar surfaces.

The discovery of quasicrystals (QCs) has changed the nature of crystallography by redefining the terminology of a crystal. QCs constitute an intermediate phase between fully periodic and fully disordered media; they lack periodicity, but exhibit long-range ordering. As a result, the diffraction pattern of QCs, consisting of sharp and discrete peaks, reveals the rotational symmetries forbidden to ordinary crystals, including five-fold symmetry. QCs were first observed via an electron diffraction from a synthesized metallic alloy [1-2], while more recently, natural QCs have been found [3]. Quasicrystalline ordering has attracted extensive attention in various realms such as architecture, mathematics, chemistry, solid-state physics, and optics [4]. The light-matter interaction in QC structures is intriguing since it is manifested by peculiar properties such as complete photonic bandgap [5], lasing [6], localization and enhanced transport [7,8], topological phase transitions [9], optical frequency conversion [10] and super-oscillations [11]. An additional twist in this field originates from the implementation of photonic QCs by metasurfaces, i.e., ultrathin two-dimensional metamaterials consisting of subwavelength meta-atoms (optical nanoantennas). Metasurfaces have triggered enormous interest as they may replace bulky optical components with ultrathin elements, paving the way for planar photonics [12-15]. Moreover, metasurfaces provide a new approach to investigate the impact of the structural symmetry on the light-matter interaction [15], which can leverage the crystallographic characteristic of complex systems. Recently, it has been shown that metallic QC metasurfaces (QCMs) consisting of isotropic nanoantennas exhibit sharp transmission resonances governed by a quasi-momentum conservation rule $\mathbf{k}_\parallel + \mathbf{G} = \mathbf{k}_{SPP}$ [16-20]. Here, $\mathbf{k}_{SPP}$ is the wavevector of surface plasmon polaritons (SPPs), i.e., surface-confined waves arising from the coupling of an electromagnetic field with the collective oscillations of quasi-free electrons at the metal surface, $\mathbf{k}_\parallel$ is the wavevector of the in-plane component of the incident illumination, and $\mathbf{G}$ is any vector from the discrete reciprocal space of the QC.

Metasurfaces can also be formed by replacing the isotropic building block by its anisotropic counterpart, e.g., rod antennas. It has been shown that periodic metasurfaces consisting of anisotropic nanoantennas with space-variant orientations $\theta(x, y)$ have such a significant effect on the structure factor that requires a correction term $\mathbf{G}_c \propto \nabla \theta(x, y)$ to the momentum-matching selection rule [15,21]. This modification was manifested by the observation of new optical modes attributed to the geometric phase pickup of $2\sigma_\pm \theta(x, y)$, acquired due to the manipulation of the polarization state of light in the anisotropic metasurface [14,22,23]. Here $\sigma_\pm = \pm 1$ denotes the polarization helicity (photon-spin in $\hbar$ units) corresponding to right and left circular polarizations, respectively. Note that this phase manifests itself by a geometric origin and it differs from a dynamic phase arising from optical path differences [24]. In periodic metasurfaces with broken inversion symmetry, the optical spin degeneracy is removed and the light-matter interaction is manifested by the optical spin-Hall effect (OSHE) [15,21,23,24], which is the optical counterpart of the spin-Hall effect in electronic transport [25]. In light of the above, it is of a great interest to investigate the influence of the local anisotropy on the light-matter interaction in QCMs, which are structures without inversion symmetry while possessing a rotational symmetry. In this letter, we report on the observation of unique discrete optical modes of a QCM as a result of an aperiodic geometric phase induced by anisotropic antennas. These modes exhibit a remarkable OSHE in QCMs when defects in the geometric phase are introduced.

Photonic QCMs were implemented in a Penrose structure by the use of standard photolithography on a SiC substrate supporting resonant collective lattice vibrations (surface phonon polaritons (SPhPs) [26]) in the infrared region. An isotropic QCM used as a reference was obtained by embedding an isotropic antenna at the middle of each edge of the Penrose tiling rhombuses [Fig. 1(a)]. Then, an anisotropic QCM was introduced by replacing the isotropic antennas with their anisotropic counterparts that were oriented along the tiling grid [Fig. 1(b)] (see also Supplemental Material Section 1 [27]). In such an arrangement, the orientation angles $\theta \in \{0°, 36°, 72°, 108°, 144°\}$ [Fig. 1(g)] correspond to the inherent anisotropy of the tiling building blocks [27]. Specifically, the QCMs were designed with the side length of the rhombuses $d = 14$ μm to match the spectral region of SiC where surface waves of SPhPs are supported. We measured the angle-resolved thermal emission spectra by a Fourier transform infrared spectrometer (FTIR) at varying polar and fixed azimuthal angles $(\vartheta, \varphi)$, respectively, while heating the samples to 773 K [see Fig. 1(g)]. The observed dispersion of the anisotropic QCM [Fig. 1(d)] reveals new collective modes as compared to the isotropic QC [Fig. 1(c)], which may give rise to geometric phase bands due to the space-variant orientation of the antennas.

To analyze the experimental results, we calculated the structure factor, i.e., the momentum space of the isotropic QCM by the Fourier transform $\hat{F}$ of the nanoantenna positions $\mathbf{r}_n$ in the real space $f_0(\mathbf{k}) = \hat{F}\left[\sum_n \delta(\mathbf{r} - \mathbf{r}_n)\right]$ [Fig. 2(c)]. The structure factor reveals a discrete wavevector set $\mathbf{G}_1$ with ten-fold rotational symmetry in the reciprocal space, associated with the characteristic distance $d_1$ in the QC structure via the relation $|\mathbf{G}_1| = 4\pi/d_1$ [16], where $d_1 = d\tau$ and $\tau = (1+\sqrt{5})/2$ is the golden ratio [Fig. 1(a)]. The measured dispersion relation of the isotropic QCM [Fig. 1(c)] exhibits good agreement with the calculation based on the momentum-matching equation $\mathbf{k}_{e\|} + \mathbf{G} = \mathbf{k}_{SPhP}$ [Fig. 1(e)]. Here, $\mathbf{k}_{e\|}$ is the wavevector component of the emitted light in the surface plane and $\mathbf{k}_{SPhP}$ is the SPhP momentum. For an analogous calculation in the case of the anisotropic QCM, the orientation degree of freedom $\theta(\mathbf{r}_n)$ must be taken into account. It was previously shown that the localized mode resonance of an anisotropic void antenna is observed with a linear polarization excitation parallel to its minor axis [23,28,29] (see also Supplemental Material Section 2 [27]). When such a dipole antenna is interacting with circularly polarized light, a local phase pickup delay of $2\sigma_\pm\theta$ is induced. Thus the momentum space of the anisotropic QCM is given by $f_\pm(\mathbf{k}) = \hat{F}\left[\sum_n \delta(\mathbf{r} - \mathbf{r}_n) e^{i2\sigma_\pm \theta(\mathbf{r}_n)}\right]$ with an additional geometric phase factor, corresponding to the orientational degree of freedom [see Fig. 2(d)]. Specifically, in the introduced anisotropic QCM, the orientation angle of the rod antennas, which is mod $\pi$ defined, is discrete due to the Penrose tiling process, i.e., $\theta \in \{0°, 36°, 72°, 108°, 144°\}$; as a result, the geometric phase is five levels quantized. Figure 2(d) shows the wavenumbers $|\mathbf{G}_m| = 4\pi/d_m$ ($m = 1,...,5$) with ten-fold rotational symmetry that are located on five circles in the momentum space, where the four inner circles ($m = 2,...,5$) reveal the new wavenumbers, whereas the outer circle of modes ($m = 1$) coincides with the isotropic structure states [compare to Fig. 2(c)]. The characteristic distances of the anisotropic QCM, depicted in Fig. 1(b), are defined via $d_{1,...,5} = d \cdot \{\tau, \gamma, \tau^2, \tau^2\gamma, \tau^3\gamma\}$, where $\gamma = \sqrt{\tau+2}$. By considering the corresponding momentum space [Fig. 2(d)], we calculated the dispersion of the anisotropic QCM according to the quasi-momentum conservation. The result [Fig. 1(f)] confirms the measured dispersion [Fig. 1(d)] and reveals the new modes in comparison to the isotropic QCM. These discrete modes arise from the geometric phase term induced by the aperiodic space-variant orientations of the antennas. Such a phenomenon is supported by observation of spectral plasmonic resonances originated from the geometric phase pickup within the visible transmission spectrum of a metal structured surface (Supplemental Material Section 3 [27]). Moreover, the polarization-resolved measurement of the thermal emission revealed that the QC modes are polarization-helicity degenerated. This result is in agreement with the calculation of the Fourier amplitudes $|f_\pm(\mathbf{k})|$ for the $\sigma_\pm$ helicities of the emerging field, and it originates from the rotational symmetry of the structures [30].

We further investigated the optical modes induced by the geometric phase in the anisotropic QCM, and specifically the polarization of light interacting with the QCM, within the visible spectrum. By reducing the length scale, we fabricated nanoscale isotropic and anisotropic QCMs, using a $Ga^+$ focused ion beam [Figs. 2(a) and 2(b)]. The samples were sandwiched between circular polarizers and normally illuminated with a continuous wave Ti:sapphire tunable laser at a wavelength of 750 nm, whereas the initial polarizer determines the incident polarization helicity of $\sigma_\pm$. The spin-projected intensity distribution was measured in the far field, which corresponds to the momentum space. The observed diffraction pattern from the isotropic QCM maintains the incident polarization [compare Figs. 2(e) and 2(g)] and confirms the single circle of modes with ten-fold symmetry and a wavenumber of $|\mathbf{G}_1|$. Polarization analysis of the anisotropic QCM shows that the diffraction pattern contains both the incident helicity state $\sigma_+$ [Fig.2 (f)] and the opposite (spin-flip) state $\sigma_-$ [Fig. 2(h)]. The $\sigma_+$ component is a signature of the structural contribution to the QCM modes with the wavenumber $|\mathbf{G}_1|$, while the spin-flip component reveals five circles of modes associated with the wavenumbers $|\mathbf{G}_{1,\dots,5}|$, corresponding to the

geometric phase contribution (Supplemental Material Section 4 [27]). Such a polarization analysis provides the route to control the diffraction of the QCMs in a spin-dependent manner. Note that the same behavior of the diffraction pattern is obtained for the incident polarization state of $\sigma_-$ [27].

An electronic spin Hall effect induced by impurities [31] inspires one to investigate the OSHE in an anisotropic QCM due to defects in the aperiodic geometric phase and to obtain spin-orbit interaction of light, i.e. spin and orbital properties become strongly coupled with each other [24]. In order to observe the OSHE in QCMs, we randomly selected nanoantennas wherein the orientations are randomly chosen from a uniform angle distribution. For a specific outcome of the randomization process, where the defects constitute half of the antennas composing the QCM, we obtained a distorted QC structure with $d = 2.5\,\mu m$ [Fig. 3(a)]. The QCM was illuminated with right and left circularly polarized light at the wavelength of $\lambda = 750\,nm$ and a spin-dependent diffraction pattern, associated with the OSHE, is observed [Figs. 3(d)-3(f)]. The OSHE in the QCM was verified by calculating the spin-dependent Fourier amplitudes [Figs. 3(b), 3(c) and 3(f)]. Moreover, the efficiency of the OSHE, i.e., $\eta_{OSHE} = (I_{\sigma_+} - I_{\sigma_-})/(I_{\sigma_+} + I_{\sigma_-})$, where $I_{\sigma_\pm}$ are the intensities of a specific mode for $\sigma_\pm$ incident helicities, respectively, is strongly affected by the concentration of the defects; it increases significantly when the concentration reaches a critical value corresponding to the mean distance between defects $a^* \leq d$ [Fig. 3(g)]. Note that the OSHE efficiency for the different modes ($\mathbf{G}_m$) of the anisotropic QCM depends on the randomization process (see Supplemental Material Section 5 [27]). Inserting defects in the geometric phase ushers in violating the rotational symmetry of the QCM that enables the spin-degeneracy removal of the optical modes.

In summary, we observed strongly affected optical modes of QCM as a result of the aperiodic geometric phase induced by anisotropic nanoantennas. The presented momentum space analysis reveals the effect of the nanoantenna orientations on the light-matter interaction in QCMs. The introduced ability to remove the spin degeneracy in QCMs by inserting defects in the aperiodic geometric phase paves the way for controlling light transport via the spin-optical aperiodic metasurfaces.

**Acknowledgements**

This research was supported by the Israel Science Foundation, the Israel Nanotechnology Focal Technology Area on Nanophotonics for Detection, and KLA-Tencor.

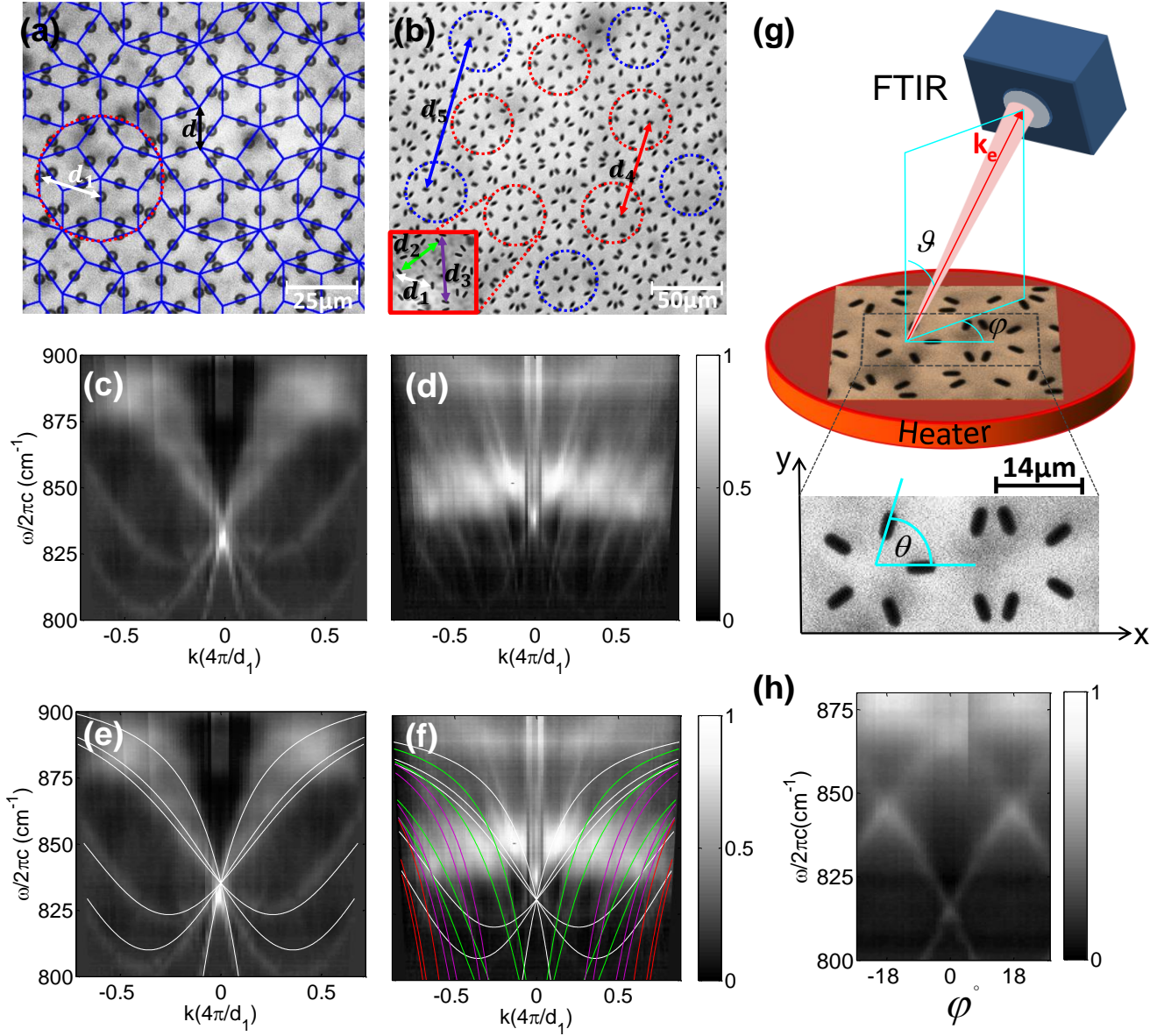

FIG. 1 (color online). Thermal emission from isotropic and anisotropic QCMs of size of 8-by-8-mm². (a) Optical microscope image of the isotropic QCM, where the blue grid depicts the Penrose tiling with a side length of the rhombuses of $d = 14$ μm. Red dashed circle represents the cluster of antennas with five-fold rotational symmetry in the real space, while the white arrow is the characteristic distance $d_1$ between the isotropic voids with a radius of 1.75 μm, etched to a depth of 1.2 μm on a SiC substrate. (b) Optical microscope image of an anisotropic QCM consisting of 1-by-4-μm² rod voids, etched to a depth of 1.2 μm on a SiC substrate. Red and blue circles depict two clusters of nanoantenas (each is rotated by $36°$ with respect to the others), where white, green and purple arrows represent characteristic distances $d_1$, $d_2$ and $d_3$ between the anisotropic antennas (see inset); $d_4$ (red) and $d_5$ (blue) are inter-cluster distances. (c,d) Dispersion $\omega(k)$ of thermal emission measured at varying polar $\vartheta \in [-50°, 50°]$ angles and a fixed azimuthal angle $\varphi = 3°$ [see Fig. 1(g)] for isotropic and anisotropic QCMs, respectively. (e,f) Momentum-matching calculation of dispersions [based on Figs. 2(c) and 2(d)] superimposed on the measurements. Red, purple and green lines in (f) highlight the new modes in the anisotropic QCM, while the white calculated modes in (e) and (f) are the signature of the isotropic structure. (g) Schematic setup for the dispersion measurement from QCMs. The thermal emission is resolved with the FTIR, $\mathbf{k}_e$ is a wavevector of the emitted light, $\vartheta$ is the emission polar angle, $\varphi$ stands for the azimuthal angle and $\theta$ is the local orientation of the anisotropic nanoantena. (h) Measured angle-resolved emission spectrum of the isotropic QCM at varying $\varphi$ and a fixed $\vartheta = 10°$. Thermal modes possessing ten-fold rotational symmetry are clearly seen.

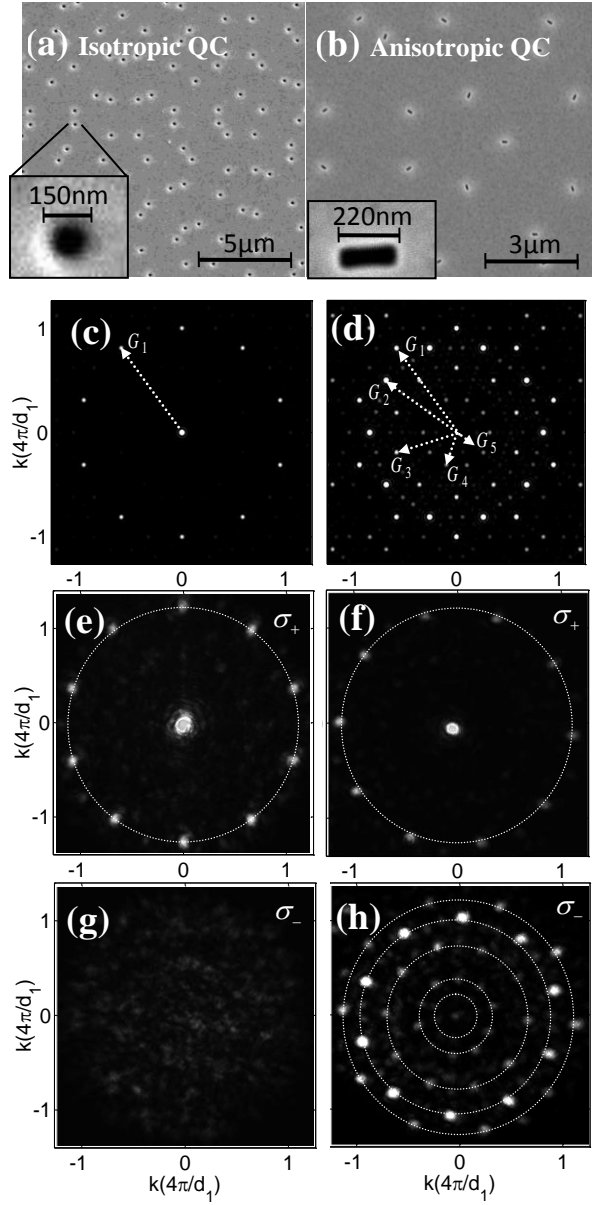

FIG. 2 (color online). (a,b) Scanning electron microscope images of isotropic and anisotropic QCMs of the size of 60-by-60-$\mu m^2$. The side of the tiling rhombi is $d = 2.5\,\mu m$, where holes of radius 60 nm (a) and 80-by-220-$nm^2$ rod apertures (b) were perforated in a 200 nm thick Au film. (c,d) Calculated diffraction intensities of the isotropic $|f_0(\mathbf{k})|^2$ (c) and anisotropic $|f_\pm(\mathbf{k})|^2$ (d) QCMs. (e,g) Measured diffraction patterns of the two helicity components of isotropic QCM: the incident helicity state $\sigma_+$ (e) and the spin-flip state $\sigma_-$ (g). (f,h) Measured diffraction patterns of anisotropic QCM of the incident helicity state $\sigma_+$ (f) and the spin-flip state $\sigma_-$ (h) components.

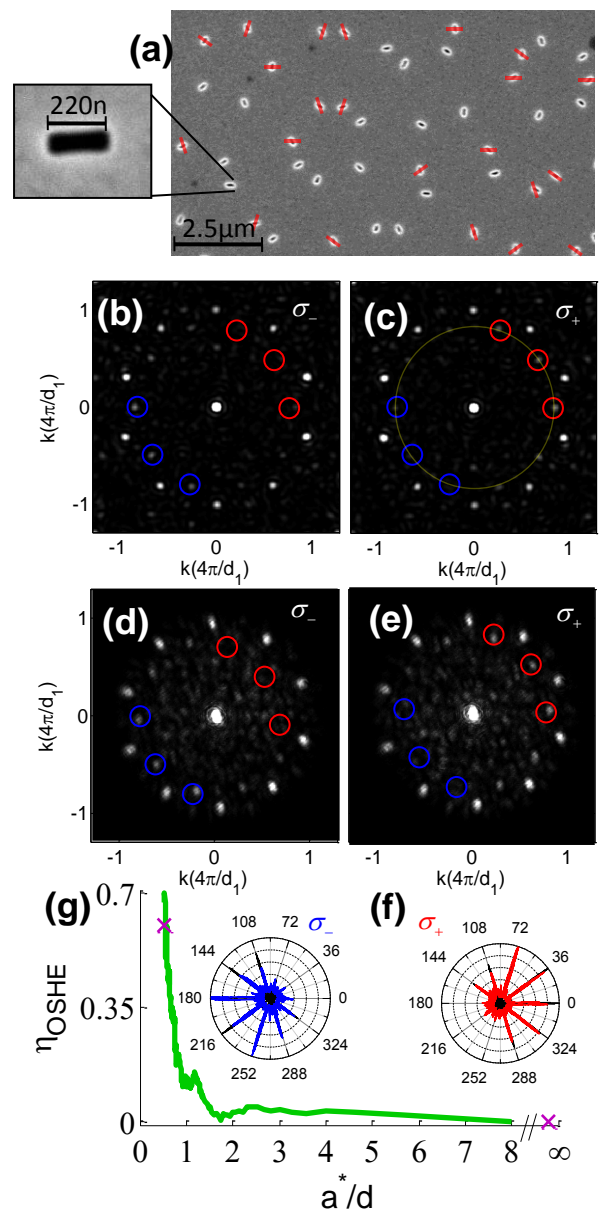

FIG. 3 (color online). (a) SEM image of the anisotropic QCM with the orientational defects. Size parameters are the same as in Fig. 2(b). Red lines denote the original antenna configuration. (b-e) Calculated (b,c) and measured (d,e) diffraction patterns of the QCM with the geometric phase defects. The patterns reveal the spin dependent modes, located on the $|\mathbf{G}_2|$ circle, for $\sigma_-$ (b,d) and $\sigma_+$ (c,e) illuminations, respectively. Blue and red guiding rings highlight the location of the spin-dependent modes in the reciprocal space for each helicity $\sigma_\pm$. (f) Azimuthal cross sections of measured (red and blue) and calculated (black) intensities for $\sigma_\pm$ illuminations. The intensities were measured along the yellow circle in (c). In this polar representation, the azimuthal angle is given in degrees and the intensity is on a linear scale. (g) Dependence of the OSHE efficiency $\eta_{OSHE}$ on the normalized average distance between defects $a^*/d$. The green curve corresponds to the calculated efficiency for the specific randomization process and the experimental points are denoted by purple crosses.

# Supplemental Material:

# Optical Mode Control by Geometric Phase in Quasicrystal Metasurface

Igor Yulevich, Elhanan Maguid, Nir Shitrit, Dekel Veksler, Vladimir Kleiner, and Erez Hasman[*]

*Micro and Nanooptics Laboratory, Faculty of Mechanical Engineering, and Russell Berrie Nanotechnology Institute, Technion – Israel Institute of Technology, Haifa 32000, Israel*

[*]e-mail: mehasman@technion.ac.il

## 1. Tiling rule of antenna configurations

Below, we provide detailed illustrations of the anisotropic antenna tiling rule. Isotropic antennas were embedded at the middle of each edge of the Penrose tiling [Fig. S1(a)] to compose an isotropic quasicrystal metasurface (QCM.) Then, an anisotropic QCM was introduced by replacing the isotropic antennas with their anisotropic counterparts, which are oriented along the tiling grid [Fig. S1(b)].

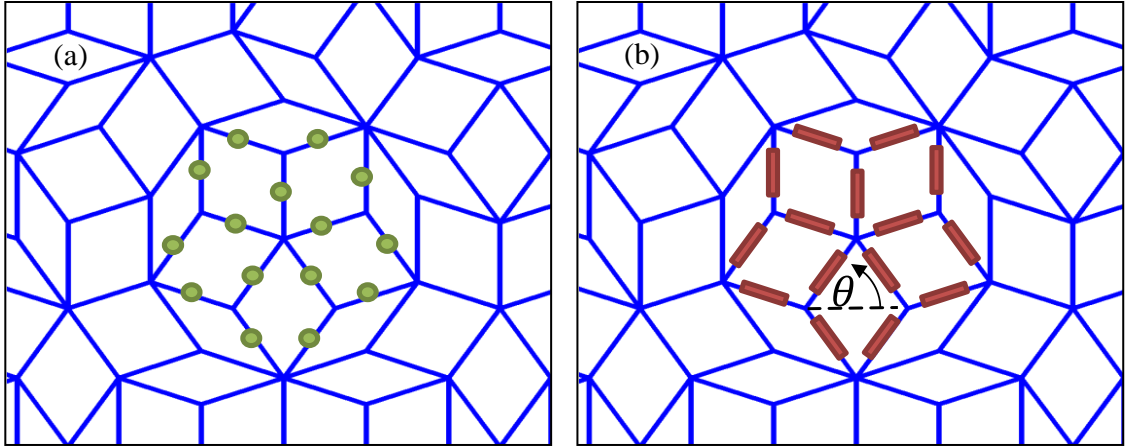

FIG. S1. Tiling rule of isotropic (a) and anisotropic (b) antenna configurations in QCM. Blue grid depicts the Penrose tiling. $\theta$ stands for the antenna orientation.

## 2. Local modes of the QCMs

In order to characterize the thermal emission from the metasurfaces, we distinguish between the collective and local mode excitations. The latter are modes arising as a result of the scattering from an individual subwavelength particle (nanoantenna) and are discussed below, while the former are modes obtained from the coherent collective scattering from nanoantennas of the QCM and governed by the momentum-matching condition discussed in the manuscript. Furthermore, these collective modes are the signature of surface phonon polaritons propagating along the SiC metasurface, thus providing such a coherent coupling of the localized modes into radiative modes.

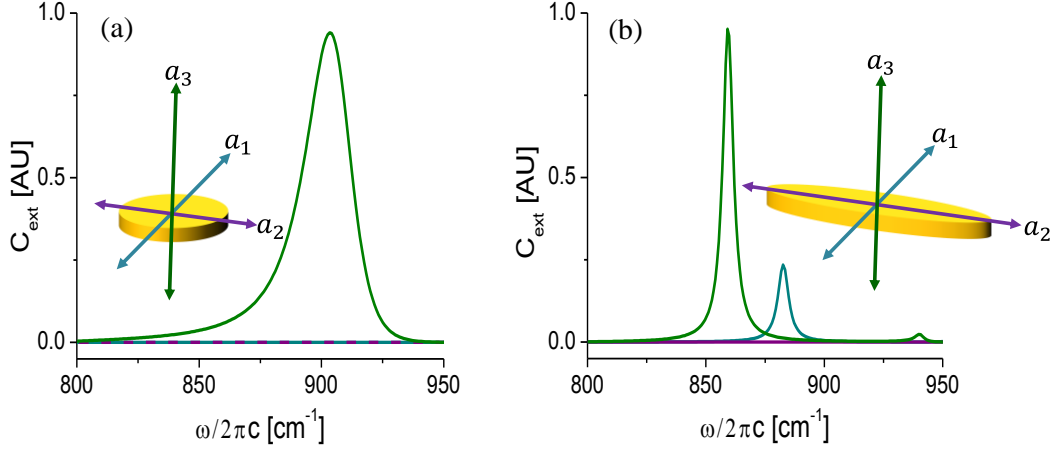

FIG. S2. (a,b) Calculated extinction cross sections of an oblate spheroid and ellipsoidal particles, respectively. The particles attained for the polarization along the semi axes $a_1$ (blue), $a_2$ (purple) and $a_3$ (green) exhibit corresponding resonances (blue, purple and green curves). The alternating blue-purple line in (a) denotes responses along both $a_1$ and $a_2$ axes.

Further, we provide an analysis of the local modes for isotropic and anisotropic antennas observed at $\omega_1/2\pi c = 885$ cm$^{-1}$ in Fig. 1(c) and at $\omega_{1,2}/2\pi c = 850$ and 887 cm$^{-1}$ in Fig. 1(d) of the manuscript. By use of the modified long wavelength approximation (MLWA) [1], we determine the extinction cross section $C_{ext} = k^4|\alpha_{0i}|^2/6\pi + k\,\text{Im}(\alpha_{0i})$ of spherically and elliptically shaped voids embedded in a SiC substrate. Here, $\alpha_{0i}$ stands for the polarizability in the MLWA regime of the $i$th semi axis of the ellipsoid, while $k$ is the normal incidence wavenumber. The correspondent polarizability is given as $\alpha_{0i} = \alpha_i\left(1 - \dfrac{k^2\alpha_i}{4\pi a_i} - \dfrac{ik^3\alpha_i}{6\pi}\right)^{-1}$, where $a_i$ ($i = 1, 2, 3$) are the semi axes of the ellipsoid. The calculated resonant frequencies $\omega_1^{MLWA}/2\pi c = 900$ cm$^{-1}$ for an oblate spheroid ($a_1 = a_2$) [Fig. S2(a)] and $\omega_1^{MLWA}/2\pi c = 859$ cm$^{-1}$ and $\omega_2^{MLWA}/2\pi c = 883$ cm$^{-1}$ for an ellipsoidal particle [Fig. S2(b)] are in good agreement with the experiment. The resonances exhibit a strong linear polarization along the direction of the small axes of the antenna [$a_3$ in Fig. S2(a) and $a_1$, $a_3$ in Fig. S2(b)], whereas for the long axes [$a_1$, $a_2$ in Fig. S2(a) and $a_2$ in Fig. S2(b)], the polarization is negligible. Such a polarization anisotropy is essential for the geometric phase accumulation in the anisotropic QCM.

### 3 (a). Transmission spectra of SPP-based QCMs

The transmission spectra of the QCMs based on nanoantennas in thin metal film were additionally studied in the visible spectrum via the excitation of SPPs. The QCMs with $d = 550$ nm, sandwiched between a glass substrate and a glass slide on the top, were normally illuminated by a supercontinuum light source (Fianium SC-450-4) and the zero-order transmitted light was collected by a spectrometer. The SPPs are resonantly excited when the momentum conservation is fulfilled; thus, the correspondent $\mathbf{k}_{SPP}$ matches the relevant circle of QC reciprocal vectors at different wavelengths. The measured transmission spectrum of the isotropic QCM [Fig. S3 top spectrum] exhibits a single resonance dip at 700 nm corresponding to the reciprocal vector set $\mathbf{G}_1$. The anisotropic QC configuration reveals an additional resonant dip at a longer wavelength of 810 nm corresponding to the reciprocal vector set $\mathbf{G}_2$ [Fig. S3 bottom spectrum]. Note that this result purely arises from the geometric phase (See section 3(b)). The obtained dip wavelengths are in good agreement with the theoretical values $\lambda_{1,2} = 2\pi\sqrt{\varepsilon_1\varepsilon_2/(\varepsilon_1+\varepsilon_2)}/|\mathbf{G}_{1,2}| = 690$ and 810 nm, where $\varepsilon_{1,2}$ stand for the glass and gold permittivity, respectively. Moreover, the inset of Fig. S3 shows the calculated finite difference time domain (FDTD) transmission spectra confirming the experimental results. The three additional modes with smaller wavevectors $\mathbf{G}_{3,4,5}$ cannot be detected within the restricted experimental spectral window and for this reason do not appear in Fig. S3.

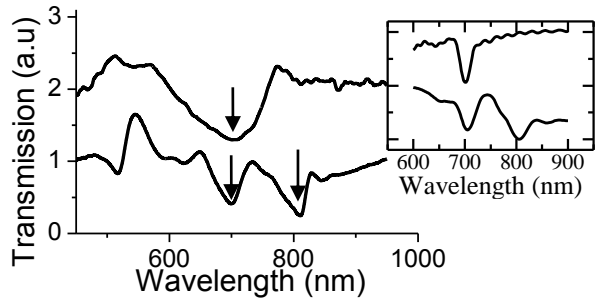

FIG. S3. Measured and calculated (inset) transmission spectra through the isotropic (top curve) and the anisotropic (bottom curve) SPP-based QCMs with extent of 40-by-40-$\mu m^2$. Apertures with the same dimensions as in Figs. 2(a) and 2(b) were perforated in a 250 nm thick Au film.

**3(b). Comparison between modes of homogenous anisotropic and inhomogeneous anisotropic QCMs**

We performed finite difference time domain (FDTD) simulations to compare the modes of the homogenous and inhomogenous orientation of nanoantennas in the anisotropic QCMs. The modes were evaluated by examining the antiresonance dips in the reflection spectrum. These dips result from the absorption to surface plasmon polaritons waves. Two quasicrystal structures having the same nanoantennas positions and different sets of orientations were simulated. The nanoantennas in the first structure were oriented along the tiling grid [see Fig. S4(b)], while the nanoantennas in the second structure have uniform orientation [see Fig. S4(a)]. Normal illumination with circularly polarized light on the inhomogeneous QCM reveals a resonance dip at

~700 nm, corresponding to the reciprocal vectors of the QCM, and two additional resonant dips at longer wavelengths of 810 nm and 950 nm, that correspond to the aperiodic orientations of the antennas [see black curve in Fig. S4(c)]. We conducted reflection simulations to find the spectrum of homogenous QCM, which is illuminated by circularly polarized light. Such an illumination reveals a broadband absorption (~150 nm), which is a signature of the local resonance of the nanoantennas and is not related to the collective modes [Fig. S4(c) blue curve].

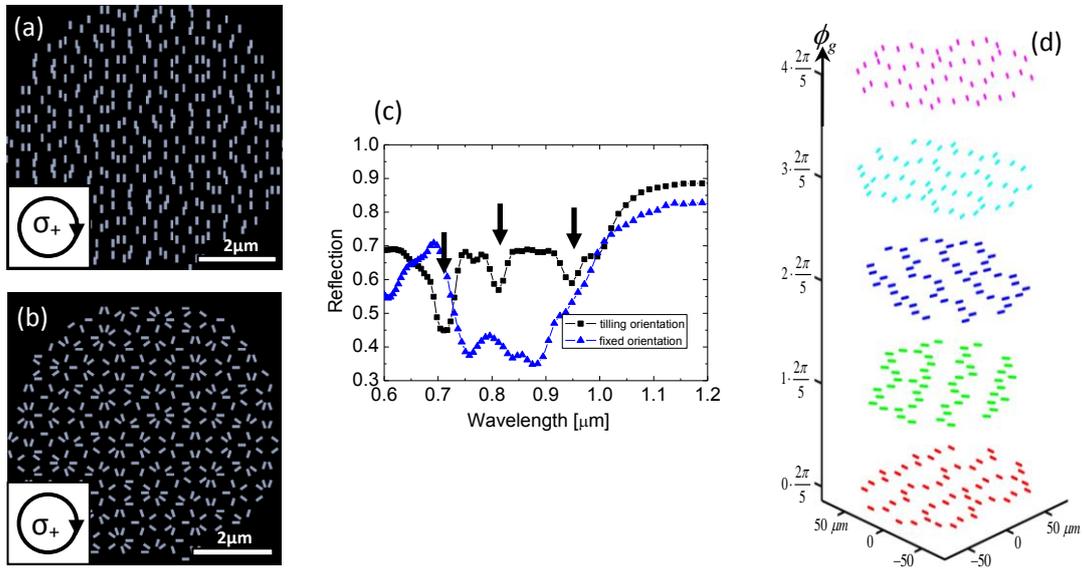

FIG. S4. Simulated reflection spectra of homogenous anisotropic QCM and inhomogeneous anisotropic QCM. (a,b) The simulated quasicrystal structures with nanoantennas oriented along the tiling grid (b) and in fixed orientation (a). The inset depicts the illuminated light polarization of each configuration. (c) The simulated reflection. The black arrows correspond to the absorption modes of the inhomogeneous QCM, while the blue curve denotes the reflection spectrum of the homogeneous QCM. (d) Five-level quantized geometric phase of the inhomogeneous

anisotropic QCM. Each color level corresponds to the same phase pickup (attached value at the left) from the equally oriented antennas.

**4. Far-field radiation from the anisotropic QCM**

In order to characterize the scattering field from the anisotropic QCM, we consider a single anisotropic antenna as a radiating dipole $\mathbf{p}_\theta = (p_{\theta x}, p_{\theta y})$ located at $\mathbf{r}$ in the $x$-$y$ interface between dielectric ($z > 0$) and polar ($z < 0$) media and oriented with an angle $\theta$ with respect to $x$ axis (see Fig. S5).

It has been shown that the scattering field in the radiation zone can be written as [2]

$$\mathbf{E} \propto \frac{e^{i\mathbf{k}\cdot(\mathbf{R}-\mathbf{r})}}{|\mathbf{R}-\mathbf{r}|}[(\hat{\mathbf{m}} \times \mathbf{p}_\theta) \times \hat{\mathbf{m}}], \tag{S1}$$

where $|\mathbf{k}| = \frac{2\pi}{\lambda}$, $\hat{\mathbf{m}}$ is the unit vector in the direction of the point of observation D and $\mathbf{R}$ is the vector directing to this point from the origin O. By assuming that $kR \gg 1$, we obtain that $\hat{\mathbf{m}} \approx \hat{\mathbf{z}}$, thus Eq. S1 takes the form

$$\mathbf{E} \propto \frac{e^{i\mathbf{k}\cdot\mathbf{R}}}{|\mathbf{R}|} e^{-i\mathbf{k}\cdot\mathbf{r}} \mathbf{p}_\theta. \tag{S2}$$

We express the dipole moment of an anisotropic antenna exited by an incident light $\mathbf{E}_{in}$ via the polarizability tensor $\ddot{\alpha}_\theta$ as

$$\mathbf{p}_\theta = \ddot{\alpha}_\theta \mathbf{E}_{in}. \tag{S3}$$

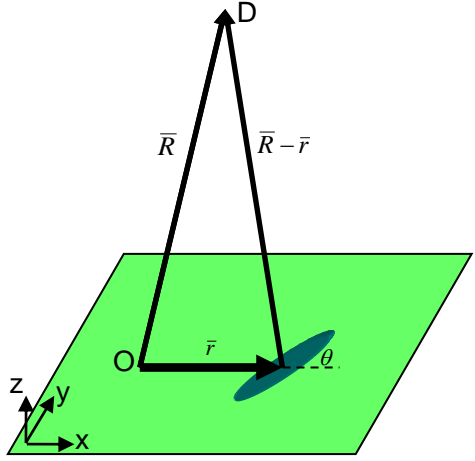

FIG. S5. Schematic setup of the radiating dipole.

The polarizability tensor of the nanoantenna, oriented with its long axis parallel to the $x$ direction, can be written as

$$\vec{\alpha}_0 = \alpha \begin{pmatrix} 1 & 0 \\ 0 & 0 \end{pmatrix}, \tag{S4}$$

while the polarizability tensor of the nanoantenna oriented at an angle $\theta$ is given by

$$\vec{\alpha}_\theta = \alpha \begin{pmatrix} \cos\theta & -\sin\theta \\ \sin\theta & \cos\theta \end{pmatrix} \begin{pmatrix} 1 & 0 \\ 0 & 0 \end{pmatrix} \begin{pmatrix} \cos\theta & \sin\theta \\ -\sin\theta & \cos\theta \end{pmatrix}. \tag{S5}$$

This transformation yields that

$$\vec{\alpha}_\theta = \frac{\alpha}{2} \begin{pmatrix} 1 & 0 \\ 0 & 1 \end{pmatrix} + \frac{\alpha}{2} \begin{pmatrix} \cos 2\theta & \sin 2\theta \\ \sin 2\theta & -\cos 2\theta \end{pmatrix}. \tag{S6}$$

For convenience, we adopt the Dirac bra-ket notation and convert $\vec{\alpha}_\theta$ to the helicity basis in which $|\sigma_+\rangle = \begin{pmatrix} 1 \\ 0 \end{pmatrix}$ and $|\sigma_-\rangle = \begin{pmatrix} 0 \\ 1 \end{pmatrix}$ are the two-dimensional unit vectors for right-handed and left-handed circular polarizations, respectively. In this

basis, the polarizability is described by the matrix $\vec{\alpha}_\theta^h = \mathbf{U}\vec{\alpha}_\theta \mathbf{U}^{-1}$, where $\mathbf{U} = \frac{1}{\sqrt{2}}\begin{pmatrix} 1 & i \\ 1 & -i \end{pmatrix}$ is a unitary conversion matrix. The explicit calculation yields that

$$\vec{\alpha}_\theta^h = \frac{\alpha}{2}\begin{pmatrix} 1 & 0 \\ 0 & 1 \end{pmatrix} + \frac{\alpha}{2}\begin{pmatrix} 0 & e^{i2\theta} \\ e^{-i2\theta} & 0 \end{pmatrix}. \tag{S7}$$

Thus, for an incident plane wave with an arbitrary polarization $|E_{in}\rangle$, we find that the resulting field is

$$|E_{out}\rangle \propto \frac{e^{i\mathbf{k}\cdot\mathbf{R}}}{|\mathbf{R}|} e^{-i\mathbf{k}\cdot\mathbf{r}} \left( |E_{in}\rangle + e^{i2\theta}\langle E_{in}|\sigma_-\rangle|\sigma_+\rangle + e^{-i2\theta}\langle E_{in}|\sigma_+\rangle|\sigma_-\rangle \right). \tag{S8}$$

The contribution from all nanoantennas located at $\mathbf{r}_n$ and with corresponding orientation angles $\theta(\mathbf{r}_n)$ in the anisotropic QCM results in the total field distribution which is given by

$$|E_{out}^{total}(\mathbf{k})\rangle \propto \frac{e^{i\mathbf{k}\cdot\mathbf{R}}}{|\mathbf{R}|} \sum_n e^{-i\mathbf{k}\cdot\mathbf{r}_n} \left( |E_{in}\rangle + e^{i2\theta(\mathbf{r}_n)}\langle E_{in}|\sigma_-\rangle|\sigma_+\rangle + e^{-i2\theta(\mathbf{r}_n)}\langle E_{in}|\sigma_+\rangle|\sigma_-\rangle \right),$$

(S9)

or alternatively,

$$|E_{out}^{total}(\mathbf{k})\rangle \propto f_0|E_{in}\rangle + f_+\langle E_{in}|\sigma_-\rangle|\sigma_+\rangle + f_-\langle E_{in}|\sigma_+\rangle|\sigma_-\rangle. \tag{S10}$$

Here, $f_0 = \hat{F}\left\{\sum_n \delta(\mathbf{r}-\mathbf{r}_n)\right\}$ and $f_\pm = \hat{F}\left\{\sum_n \delta(\mathbf{r}-\mathbf{r}_n)e^{\pm i2\theta(\mathbf{r}_n)}\right\}$. Equation S10 shows that $|E_{out}^{total}(\mathbf{k})\rangle$ comprises three polarization orders: the $|E_{in}\rangle$, $|\sigma_+\rangle$ and $|\sigma_-\rangle$. The $|E_{in}\rangle$ polarization order maintains the polarization and phase of the incident beam, whereas the phases of $|\sigma_\pm\rangle$ polarization orders is equal to $\pm 2\theta$, respectively. Note that the phase modification of the $|\sigma_+\rangle$ and $|\sigma_-\rangle$ polarization orders results solely from local changes in polarization and is therefore, geometric in nature.

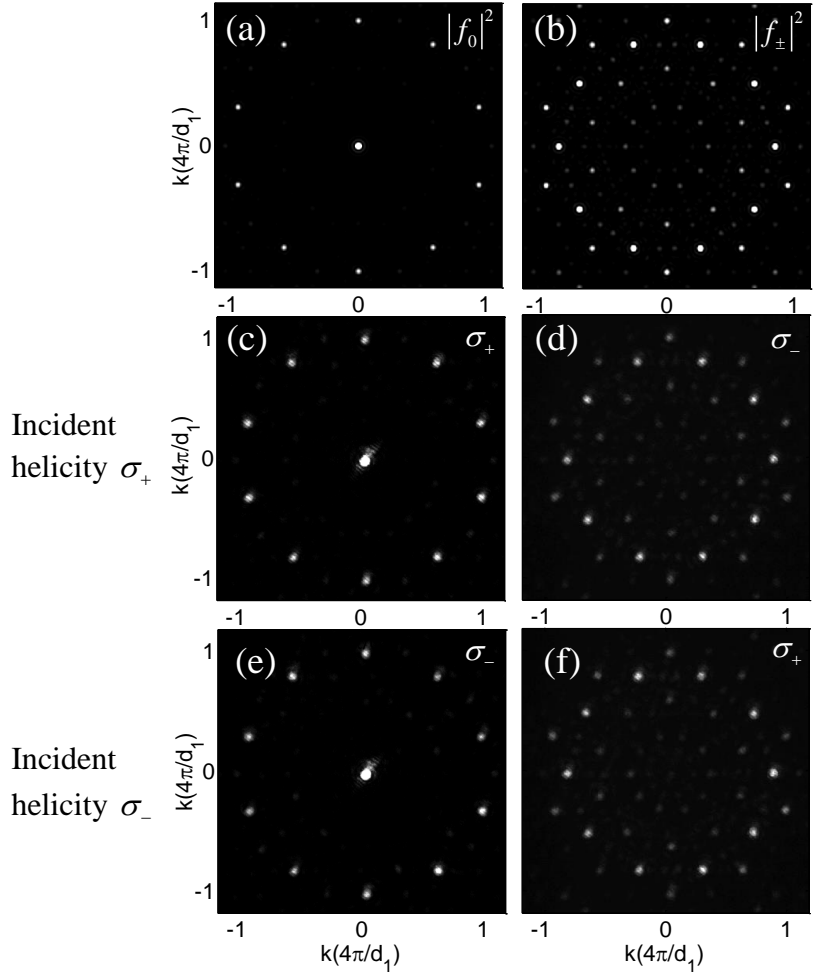

FIG. S6. (a,b) Calculated $|f_0|^2$ (a) and $|f_\pm|^2$ (b) corresponding to the first term (incident helicity component) and the second term (spin-flip component) of Eqs. S11 and S12. (c-f) Measured diffraction patterns of the two helicity components of the anisotropic QCM for incident spin state $|\sigma_+\rangle$ (c,d) and incident spin state $|\sigma_-\rangle$ (e,f). Measured diffraction patterns of the incident helicity component (c,e) and the spin-flip component (d,f). $\sigma_\pm$ in the figures denote the spin-projected states of the measured diffracted waves. Note that the examined QCM was the same as presented in Fig. 2(b) with the same experimental parameters.

For the incident spin state of $|\sigma_+\rangle$, Eq. S10 can be expressed as

$$\left|E_{total}^{\sigma_+}(\mathbf{k})\right\rangle \propto f_0|\sigma_+\rangle + f_-|\sigma_-\rangle, \tag{S11}$$

while for the incident spin state of $|\sigma_-\rangle$, Eq. S10 is given by

$$\left|E_{total}^{\sigma_-}(\mathbf{k})\right\rangle \propto f_0|\sigma_-\rangle + f_+|\sigma_+\rangle. \tag{S12}$$

The presented model exhibits a good agreement with the observation of the field scattered from the anisotropic QCMs. The first term of Eqs. S11 and S12 ushers in the diffraction pattern [Fig. S6(a)] of the field with the incident helicity component, and it confirms the measured single circle of modes [Fig. 2(f) and Figs. S6(c) and S6(e)]. The emerging spin-flip component leads to the calculated diffraction pattern [Fig. S6(b)] which is the signature of the geometric phase modes manifested by the five circles, confirming the measured diffractions for the incident spin states of $|\sigma_+\rangle$ [Fig. 2(h) and Fig. S6(d)] and $|\sigma_-\rangle$ [Fig. S6(f)].

**5. Optical spin-Hall effect for different outcomes of the randomization process**

For a random selection of half of the anisotropic nanoantennas in the QCMs, we obtained a part of the QCM for which the orientations would be distorted. By the repeated random process of the antenna orientations configuration, we achieved an outcome, where spin-dependent modes are obtained on the circle of radius $|\mathbf{G}_3|$ [Figs. S7(a), S7(b) and S7(e)]. This spin-dependent diffraction pattern demonstrates the optical spin-Hall effect (OSHE) in the perturbed anisotropic QCM. The QCM was illuminated with right and left circularly polarized light at the wavelength of $\lambda = 750$ nm and the OSHE was observed experimentally [Figs. S7(c)-S7(e)]. The efficiency of the OSHE becomes significant when half of the antenna defects are introduced in the structure [Fig. S7 (f)].

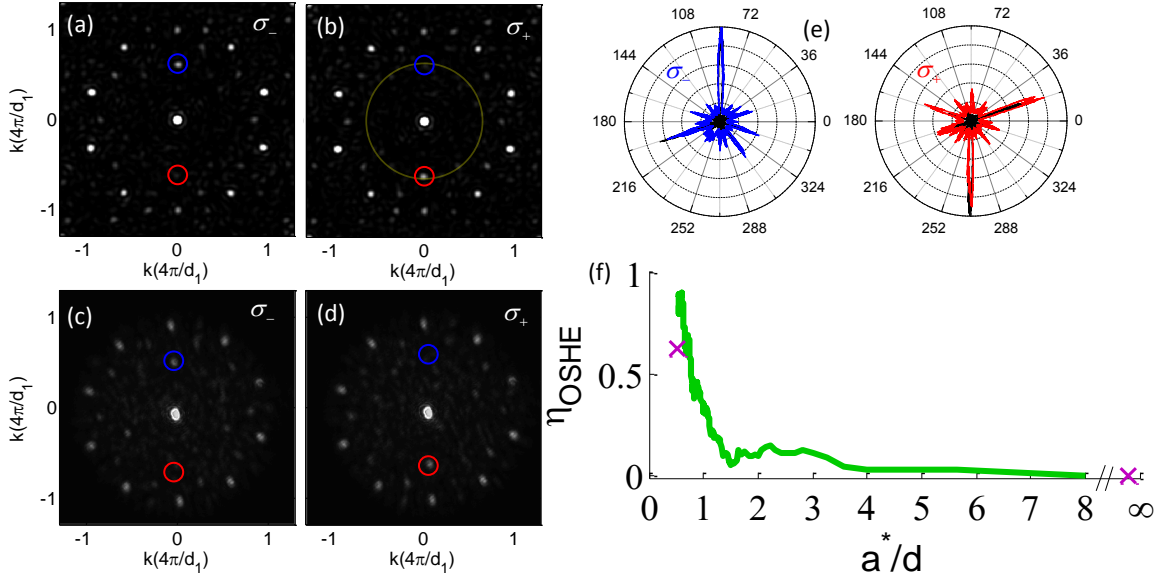

FIG. S7. (a-d) Calculated (a,b) and measured (c,d) diffraction patterns of the QCM with the geometric phase defects. The patterns reveal the spin-dependent modes, located on the $|\mathbf{G}_3|$ circle, for $\sigma_-$ (a,c) and $\sigma_+$ (b,d) illuminations, respectively. Blue and red guiding rings highlight the location of the spin-dependent modes in the reciprocal space for each helicity $\sigma_\pm$. (e) Azimuthal cross sections of measured (red and blue) and calculated (black) intensities for $\sigma_\pm$ illuminations. The intensities were measured along the yellow circle in (b). In this polar representation, the azimuthal angle is given in degrees and the intensity is on a linear scale. (f) Dependence of the OSHE efficiency $\eta_{OSHE}$ on the normalized average distance between defects $a^*/d$. The green curve corresponds to the calculated efficiency for the specific randomization process and the experimental points are denoted by purple crosses.